\documentclass[prd, preprint, nofootinbib]{revtex4}

\usepackage{graphicx}
\usepackage{amsfonts}
\usepackage{latexsym}
\usepackage{amssymb}
\usepackage{epsfig}

\begin{document}

\title{ 
Galactic Center gamma-ray excess \\ 
from two-Higgs-doublet-portal dark matter
}

\author{Nobuchika Okada}
 \email{okadan@ua.edu}
 \affiliation{
Department of Physics and Astronomy, 
University of Alabama, Tuscaloosa, Alabama 35487, USA
}

\author{Osamu Seto}
 \email{seto@phyics.umn.edu}
 \affiliation{
 Department of Life Science and Technology,
 Hokkai-Gakuen University,
 Sapporo 062-8605, Japan
}

%

\begin{abstract}
We consider a simple extension of the type-II two-Higgs-doublet model 
  by introducing a real scalar as a candidate for dark matter in the present Universe. 
The main annihilation mode of the dark matter particle with a mass of around $31-40$ GeV is 
  into a $b\bar{b}$ pair, and this annihilation mode suitably explains the observed excess 
  of the gamma-ray flux from the Galactic Center. 
We identify the parameter region of the model that can fit the gamma-ray excess and
  satisfy phenomenological constraints, such as the observed dark matter relic density and the null results 
  of direct dark matter search experiments. 
Most of the parameter region is found to be within the search reach of 
  future direct dark matter detection experiments. 
\end{abstract}


\preprint{HGU-CAP 036} 

\vspace*{3cm}
\maketitle


\section{Introduction}

The weakly interacting massive particle (WIMP) is a primary candidate for dark matter (DM) in the present Universe, 
   and one of the major topics in particle physics and cosmology is to reveal the nature of WIMP dark matter. 
There are many current experiments aimed at the direct or indirect detection of DM. 

Over the past several years, many analyses have shown an excess of gamma rays from the Galactic Center, 
   and interpretations involving the annihilation of DM particles~\cite{Goodenough:2009gk,Hooper:2010mq, Hooper:2011ti,Abazajian:2012pn, 
   Abazajian:2014fta} have been considered to explain this excess. 
Similarly, an excess of gamma-rays from the so-called Fermi bubble region~\cite{Su:2010qj} 
   found in the Fermi-LAT data has been interpreted as a result of indirect dark matter particle 
   detection~\cite{Hooper:2013rwa,Huang:2013pda}.

Previous studies of this gamma-ray excess have shown that the excess can be fit by a DM particle 
   with a mass of around $10$ GeV annihilating into a pair of tau leptons, or by 
   a dark matter particle with a mass of $30-60$ GeV annihilating into 
   a $b \bar{b}$ pair~\cite{Hooper:2010mq,Abazajian:2012pn,Hooper:2013rwa,Huang:2013pda,Abazajian:2014fta}. 
In addition, gamma-ray data from subhalos also show a similar spectrum shape, 
   indicating that those observations may originate from such the dark matter particles as well~\cite{Berlin:2013dva}. 
Interestingly, the DM annihilation cross section that fits the data is found to be of the same order 
   as a typical thermal annihilation cross section, $\sigma v \simeq 3 \times 10^{-26}\, {\rm cm^3/s}$, 
   for WIMP dark matter. 
Particle physics models have been proposed that could naturally realize such DM particles; 
 see, for example, Refs.~\cite{Hagiwara:2013qya, Kyae:2013qna,Okada:2013bna,Huang:2013apa,Modak:2013jya} 
   for light DM models in which a pair of DM particles annihilates into tau leptons.

However, a more recent analysis~\cite{Daylan:2014rsa} has claimed that a dark matter particle 
   with a mass of $31-40$ GeV provides an excellent fit for the gamma-ray excess, where the main annihilation 
   mode is into $b\bar{b}$ and the cross section is $ \sigma v = (1.4-2.0) \times 10^{-26} $ cm${}^3/$s. 
Fit using the annihilation mode into tau lepton pairs are no longer favored~\cite{Daylan:2014rsa}, 
   and the cross section of the tau lepton annihilation mode is severely constrained by 
   cosmic-ray positron data~\cite{Bringmann:2014lpa} (see, however, Ref.~\cite{Lacroix:2014eea}). 
Although a certain astrophysical source might be able to explain the excess~\cite{Yuan:2014rca,Carlson:2014cwa},
   the interpretation involving annihilating DM particles is a very interesting possibility and 
   various particle models employing this mechanism have been recently proposed.  
In the context of supersymmetric models, neutralino DM 
   in the next-to-minimal supersymmetric Standard Model (NMSSM)~\cite{Berlin:2014pya,Cheung:2014lqa,Huang:2014cla},
   the sneutrino~\cite{Cerdeno:2014cda} in the seesaw-extended NMSSM~\cite{Cerdeno:2008ep},
   and the sneutrino in the supersymmetric inverse seesaw model~\cite{Ghosh:2014pwa} all
   play the role of the DM.   
For nonsupersymmetric DM models, see, for example, 
   Refs.~\cite{Boehm:2014hva,Agrawal:2014una,Izaguirre:2014vva,Ipek:2014gua,Ko:2014gha,Abdullah:2014lla,Martin:2014sxa,Basak:2014sza,Cline:2014dwa,Wang:2014elb,Arina:2014yna,Ko:2014loa}.

In this paper, we propose a model in the class of so-called Higgs-portal DM models 
   to explain the gamma-ray excess, where a real scalar singlet $\phi$  
   under the SM gauge groups is introduced as a dark matter candidate, 
   along with a $Z_2$ parity that ensures the stability of the scalar. 
In the simplest model, the real scalar is a unique field that is added to the SM particle content
   (for an incomplete list, see,  e.g., Refs.~\cite{McDonald:1993ex,Burgess:2000yq, Davoudiasl:2004be,Kikuchi:2007az,KMNO}). 
However, this minimal model is not suitable for explaining the gamma-ray excess, as the desired DM mass range 
    of $31-40$ GeV is excluded by the null results of direct DM search experiments 
    (see, for example, Refs.~\cite{KMNO, deSimone:2014pda}).
Thus, we extend the Higgs sector to the two-Higgs-doublet model~\cite{Aoki:2009pf,Goh:2009wg,Cai:2011kb,Boucenna:2011hy}.
In fact, in our previous work~\cite{Okada:2013bna} we considered a Higgs-portal DM 
   in the context of the type-X two-Higgs-doublet model, where a pair of DM particles mainly annihilates into tau leptons.~\footnote{
   This part of the Higgs sector is motivated by a radiative generation of neutrino masses~\cite{Aoki:2008av}.}
Motivated by the recent analysis in Ref.~\cite{Daylan:2014rsa},
  in this paper we propose a Higgs-portal DM 
   with a mass of $31-40$ GeV in the context of the type-II two-Higgs-doublet model.  
In this case, a pair of scalar DM particles mainly annihilates to a $b\bar{b}$ pair 
   through the $s$-channel exchange of Higgs bosons with type-II Yukawa couplings. 
We will identify a model parameter region that not only explains the gamma-ray excess, 
   but that is also consistent with phenomenological constraints, such as the observed DM relic abundance 
   and the null results of the current direct DM search experiments. 
In addition, we will see that most of the identified parameter region can be covered 
  by the search reach of future direct DM detection experiments.

\section{Two-Higgs-doublet-portal scalar dark matter}

We introduce a real SM gauge singlet scalar $\phi$ as the dark matter candidate, 
   along with the $Z_2$ parity by which the stability of the DM particle is guaranteed. 
The Higgs sector is extended to the so-called type-II two-Higgs-doublet model,
   where one Higgs doublet generates the mass of the SM up-type fermions 
   while the other generates the mass of the SM down-type fermions, just as in the MSSM.
In the type-II model, the Yukawa interaction is given by 
\begin{eqnarray}
 {\cal L}_Y 
   = - y_{\ell_i}  \overline{L}^i \Phi_1 \ell_R^i
     - y_{u_i}  \overline{Q}^i \tilde{\Phi}_2 u_R^i
     - y_{d_i}  \overline{Q}^i \Phi_1 d_R^i + {\rm H.c.}, 
 \label{eq:yukawa1}
\end{eqnarray}
where $Q^i$ ($L^i$) is the ordinary left-handed SU(2) doublet quark (lepton) of the $i$th generation, and 
   $u_R^i$, $d_R^i$ and $e_R^i$ are the right-handed SU(2) singlet up- and down-type quarks 
   and charged leptons, respectively.   
Here, we have neglected the flavor mixing for simplicity.

The scalar potential for the two Higgs doublets ($\Phi_1$ and $\Phi_2$) and the scalar DM is given by
\begin{eqnarray}
 V &=& -\mu_1^2 |\Phi_1|^2 -\mu_2^2 |\Phi_2|^2 - (\mu_{12}^2
  \Phi_1^\dagger \Phi_2 + {\rm H.c.}) \nonumber \\
 &&+ \lambda_1|\Phi_1|^4   +
  \lambda_2|\Phi_2|^4 + \lambda_3|\Phi_1|^2|\Phi_2|^2 +\lambda_4 |\Phi_1^\dagger \Phi_2|^2 \  +
   \left\{ \frac{\lambda_5}{2} (\Phi_1^\dagger
    \Phi_2)^2 + {\rm H.c.} \right\} \nonumber\\ 
 &&  + \frac{1}{2}\mu_\phi^2 \phi^2 +
  \lambda_\eta \phi^4 
   + ( \sigma_1 |\Phi_1|^2 + \sigma_2 |\Phi_2|^2 ) \frac{\phi^2}{2}.
\end{eqnarray}
The electrically neutral components of the two-Higgs doublets develop vacuum expectation values, 
 which we parametrize as
\begin{eqnarray}
  \Phi_1
   = 
  \left( \begin{array}{c}
          0 \\
          \frac{v_1 + h_1}{\sqrt{2}} \\
         \end{array}
  \right), \hspace{1cm}
%
  \Phi_2
   = 
  \left( \begin{array}{c}
          0 \\
          \frac{v_2 + h_2}{\sqrt{2}} \\
         \end{array}
  \right),
\end{eqnarray}
   where the vacuum expectation values are given by $v_1=v \cos \beta$ and $v_2=v \sin \beta$ with $v=246$ GeV. 
The physical states $h_1$ and $h_2$ are diagonalized to the mass eigenstates ($h$ and $H$) as 
 \begin{eqnarray}
  \left( \begin{array}{c}
          h_1  \\
          h_2  \\
         \end{array}\right)
   = 
  \left( \begin{array}{cc}
          \cos\alpha & - \sin\alpha\\
          \sin\alpha & \cos\alpha\\
         \end{array}\right)
  \left( \begin{array}{c}
          H \\
          h \\
         \end{array}\right) .
 \end{eqnarray}
In this paper, we consider the case that the mixing angle $\alpha$ satisfies the condition 
 $\sin(\beta-\alpha) = 1$, which is the so-called SM limit,
 so that the mass eigenstate $h$ is the SM-like Higgs boson.~\footnote{
In the following, we consider the extra Higgs boson mass to be $60-80$ GeV. 
The result of the Higgs boson search at LEP~\cite{LEP2} has severely constrained 
  the coupling of such a light extra Higgs boson to the $Z$ boson, 
  which leads to $\cos^2(\beta-\alpha) \lesssim 0.01$. 
Thus, we take the SM limit for simplicity. 
}

In terms of the mass eigenstates, the (three-point) interactions 
 of the scalar dark matter $\phi$ with the Higgs bosons ($h$ or $H$) are given by 
\begin{eqnarray}
{\cal L}_\sigma \supset 
 - \frac{\sigma_1 \cos\alpha\cos\beta+\sigma_2\sin\alpha\sin\beta}{2}
  v H \phi^2 
- \frac{- \sigma_1 \sin\alpha\cos\beta+\sigma_2\cos\alpha\sin\beta}{2}
  v h \phi^2.
\label{HiggsInt}
 \end{eqnarray}
The Yukawa interactions with quarks and leptons 
 in Eq.~(\ref{eq:yukawa1}) can then be written as
\begin{eqnarray}
 {\cal L}_Y^{\rm Quarks} &\supset&
 \frac{m_{u^i} \sin\alpha}{v \sin\beta} H \bar u^i u^i
 + \frac{m_{u^i} \cos\alpha}{v \sin\beta} h \bar u^i u^i
  + \frac{m_{d^i}\cos\alpha}{v\cos\beta} H \bar d^i d^i 
  - \frac{m_{d^i}\sin\alpha}{v\cos\beta} h \bar d^i d^i , \\
{\cal L}_Y^{\rm Leptons} &\supset &
   \frac{m_{\ell^i}}{v} \frac{\cos\alpha}{\cos\beta} H \bar \ell^i \ell^i
   - \frac{m_{\ell^i}}{v} \frac{\sin\alpha}{\cos\beta} h \bar \ell^i \ell^i.
 \end{eqnarray}
Since we have set $\sin(\beta-\alpha)= 1$, the coupling between the non-SM-like Higgs ($H$) 
   and down-type quarks (charged leptons) are enhanced for $\tan\beta > 1$,
   and those between $H$ and up-type quarks are suppressed by $1/\tan\beta$,~\footnote{
This leads to a suppression of the non-SM-like Higgs boson production through gluon fusion at the LHC. 
Even if non-SM-like Higgs bosons are produced, each $H$ mainly decays to a $b \bar{b}$ pair,  
   and this decay mode is difficult to observe at the LHC.  
} 
   while the Yukawa couplings between the SM-like Higgs boson $h$ and the SM fermions remain the same as those in the SM. 
For simplicity, we fix the other model parameters so as to make the charged and $CP$-odd Higgs bosons 
 heavy enough to be consistent with their current experimental mass bound 
 and to decouple them from our analysis of the dark matter physics.

\begin{figure}[h,t]
\begin{center}
\epsfig{file=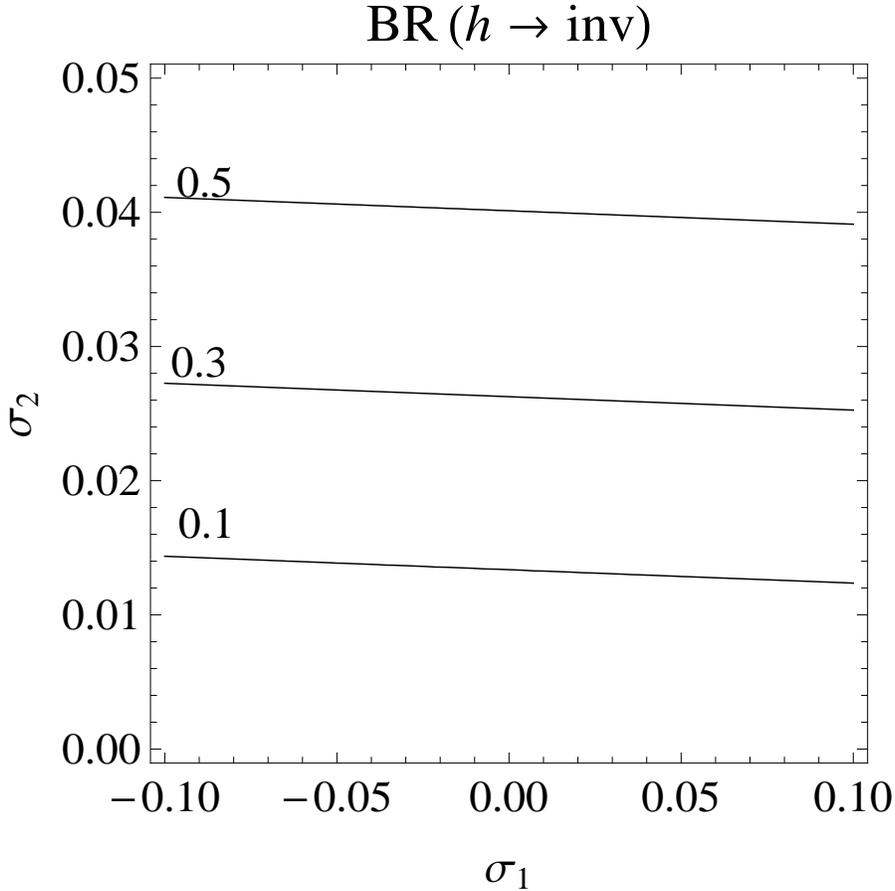, width=12cm,height=12cm,angle=0}
\end{center}
\caption{
Contours of the invisible decay branching ratio of the SM-like Higgs boson, 
  ${\rm BR}(h \rightarrow \phi \phi)=0.1$, $0.3$, and $0.5$. 
We have taken the DM mass $m_\phi=40$ GeV and $\tan \beta =10$. 
 }
\label{Fig:invisible}
\end{figure}

We first calculate the invisible decay width of the SM-like Higgs boson into a pair of
 scalar DM particles through the interactions in Eq.~(\ref{HiggsInt}).\footnote{
As we will see in the following, the non-SM-like Higgs boson $H$ is light, 
   and the SM-like Higgs boson can decay to a pair of $H$ bosons. 
To simplify our analysis, we fix the free parameters in the scalar potential to suppress this decay rate. } 
Figure~\ref{Fig:invisible} shows the branching ratio of this invisible decay ${\rm BR}(h \rightarrow \phi \phi)$ 
   for the DM mass $m_\phi=30$ GeV and $\tan \beta =10$.  
We have found that the upper bound from the LHC data,  ${\rm BR}(h \rightarrow \phi \phi) \lesssim 0.35$~\cite{Hinv}, 
   is satisfied for $\sigma_2 \lesssim 0.03$, almost independently of $\sigma_1$. 

Next, we estimate the thermal relic abundance of the real scalar DM by solving the Boltzmann equation,
\begin{equation}
 \frac{d n }{dt}+3H n =-\langle\sigma v\rangle ( n^2 - n_{\rm EQ}^2),
\end{equation}
  where $H$ and $n_{\rm EQ}$ are the Hubble parameter and the DM number density at thermal equilibrium, 
  respectively~\cite{KolbTurner}. 
The resultant thermal relic abundance is approximated as 
\begin{equation}
\Omega_{\rm DM} h^2 =
 \frac{1.1 \times 10^9 (m_\phi/T_d) \; {\rm GeV^{-1}}}
 {\sqrt{g_*}M_P\langle\sigma v\rangle}, 
\end{equation}
 where $M_P=1.22 \times 10^{19}$ GeV is the Planck mass, 
 $\langle\sigma v\rangle$ is the thermal averaged product 
 of the annihilation cross section and the relative velocity, 
 $g_*$ is the total number of relativistic degrees of freedom 
 in the thermal bath,  and $T_d$ is the decoupling temperature.

The present annihilation cross section $(\sigma v)_0$ of the DM particle --
   which is relevant for the indirect detection of dark matter -- is given by its $s$-wave component 
   of the annihilation cross section, e.g.,  by the limit of $v \rightarrow 0$. 
Note that in general the thermal averaged cross section $\langle\sigma v\rangle$ determined 
    by the condition of $\Omega h^2 \simeq 0.1$~\cite{WMAP,Planck} is not the same as 
    the present annihilation cross section $(\sigma v)_0$. 
This difference becomes significant for two cases: one is when the DM annihilation cross section 
    has a sizable $p$-wave contribution, and the other is when the dark matter mass is close to 
    a resonance pole of the mediators in the annihilation process. 
In fact, the latter is the case considered here.

\begin{figure}[htbp]
 \begin{minipage}{0.32\hsize}
  \begin{center}
\includegraphics[width=50mm]{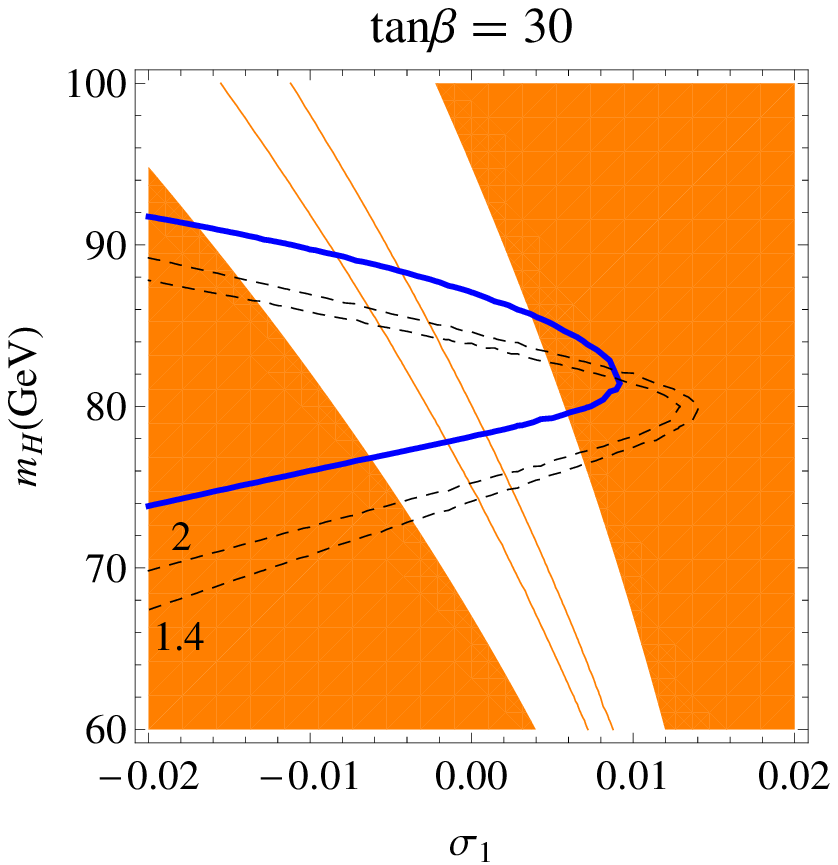}
  \end{center}
\end{minipage}
\hfill
 \begin{minipage}{0.32\hsize} 
  \begin{center}
\includegraphics[width=50mm]{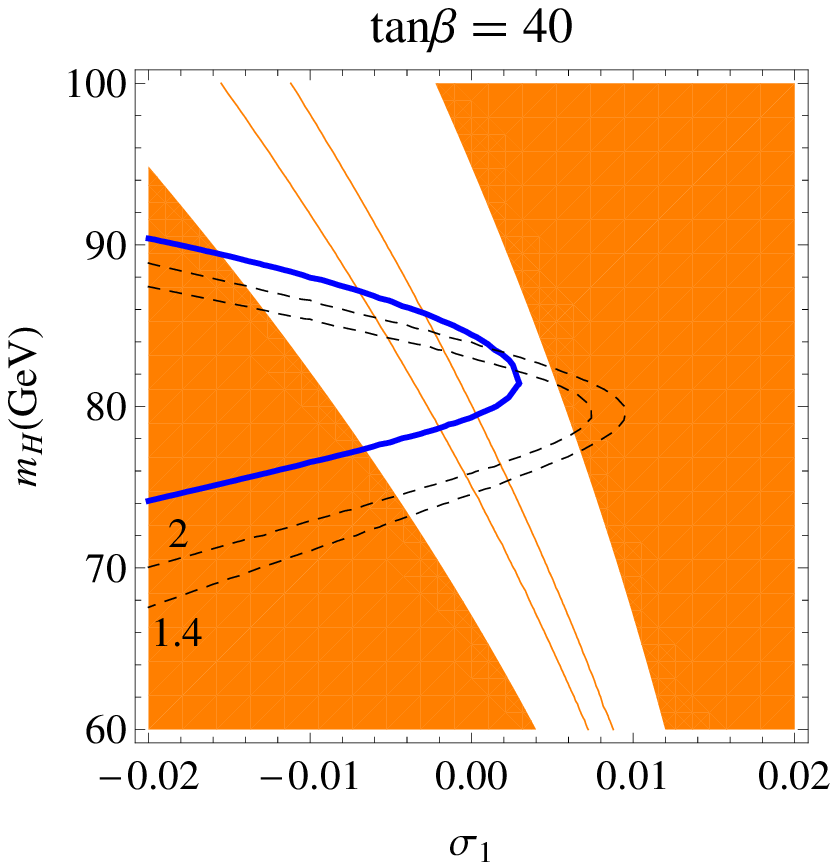}
  \end{center}
\end{minipage}
\hfill
 \begin{minipage}{0.32\hsize} 
  \begin{center}
\includegraphics[width=50mm]{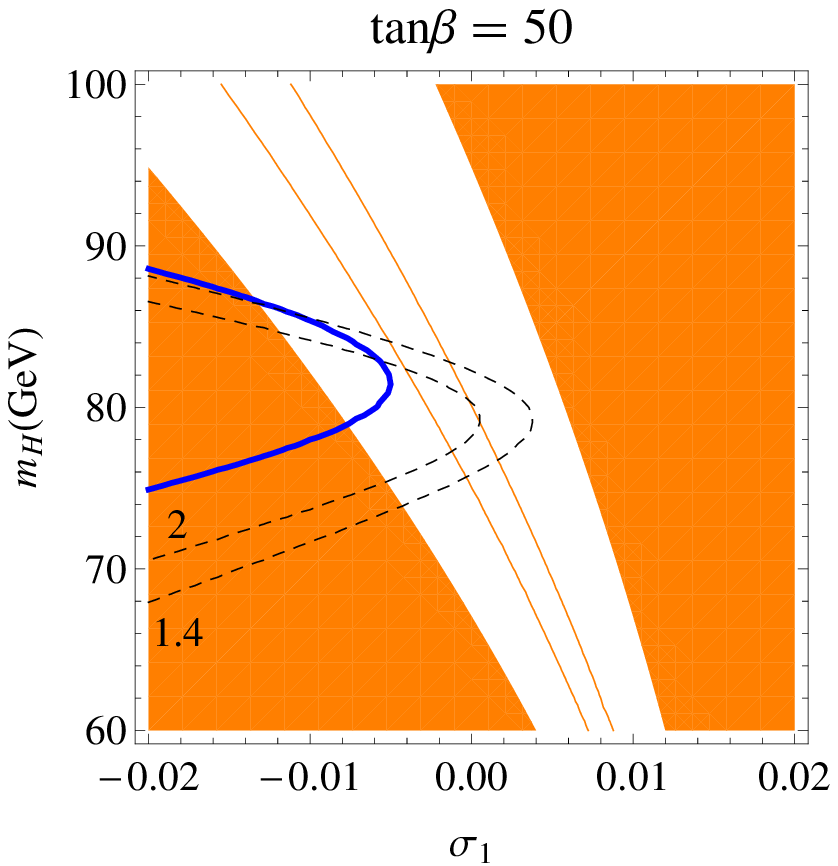}
  \end{center}
\end{minipage}
\caption{
Contours of $\Omega h^2 = 0.1$ (thick blue line) 
 and $(\sigma v)_0$ in units of  $ 10^{-26}$ cm$^3$/s 
 (dashed lines) for $m_{\rm DM}=40$ GeV and $\sigma_2=0.02$.
$\tan\beta$ is taken to be $30, 40$, and $50$ from left to right.
The shaded regions are excluded by the LUX (2014) experiment~\cite{LUX:2013}, 
 and the expected future sensitivity (  
 $3\times 10^{-47}\, {\rm cm}^2$)
 of the XENON1T experiment~\cite{XENON1T} 
 is depicted by the thin lines. 
}
\label{Fig:0240}
\end{figure}

\begin{figure}[htbp]
 \begin{minipage}{0.32\hsize}
  \begin{center}
\includegraphics[width=50mm]{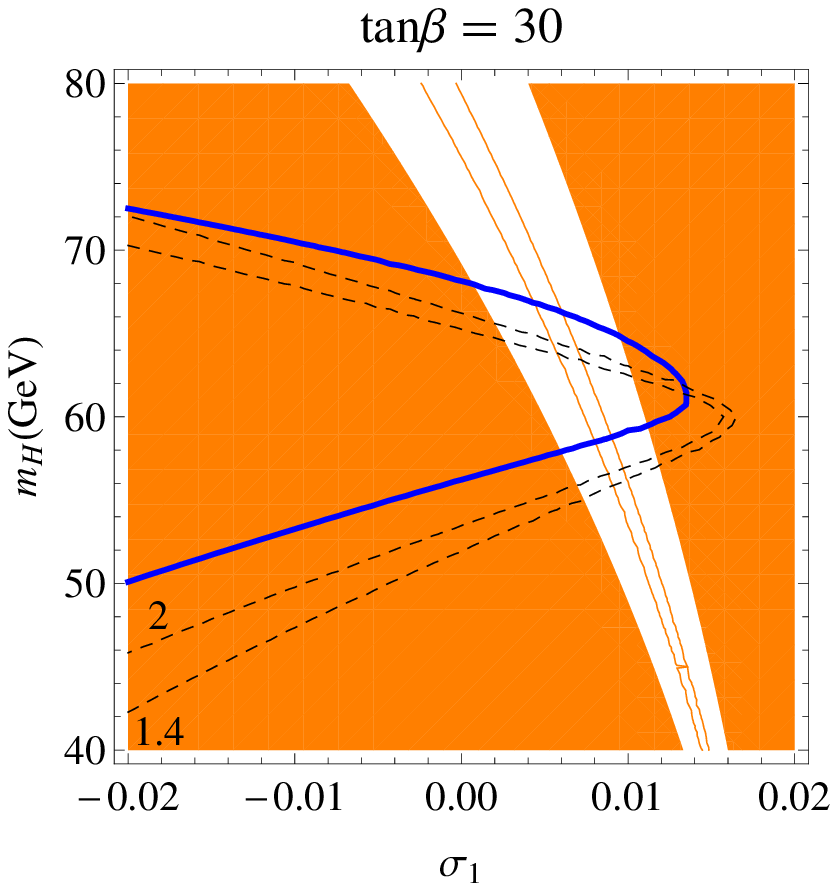}
  \end{center}
\end{minipage}
\hfill
 \begin{minipage}{0.32\hsize} 
  \begin{center}
\includegraphics[width=50mm]{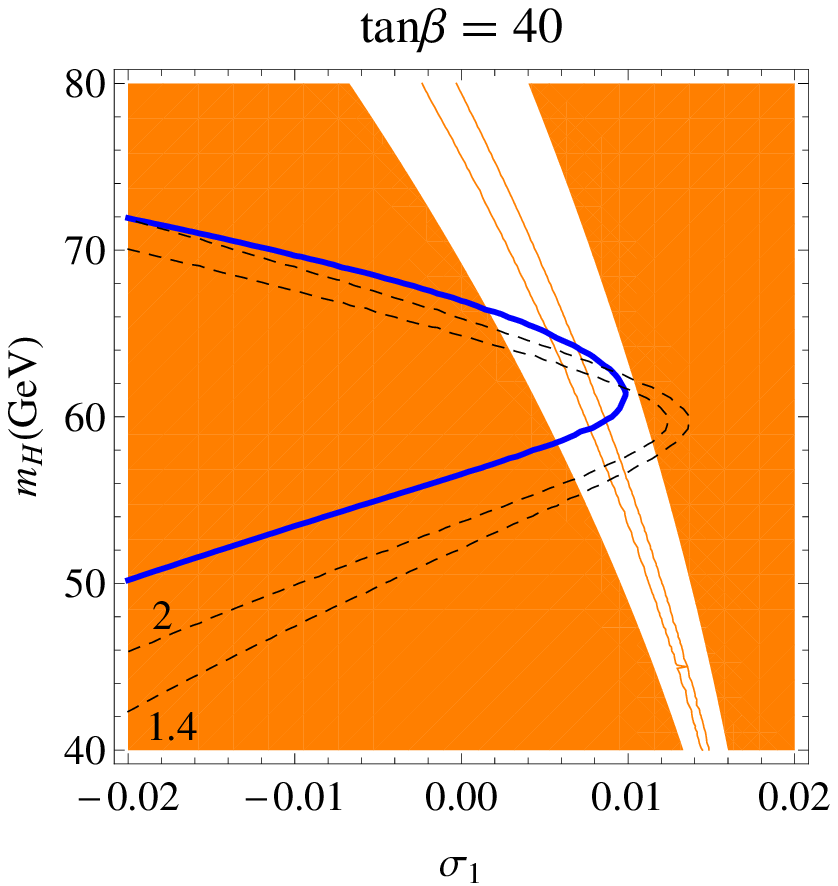}
  \end{center}
\end{minipage}
\hfill
 \begin{minipage}{0.32\hsize} 
  \begin{center}
\includegraphics[width=50mm]{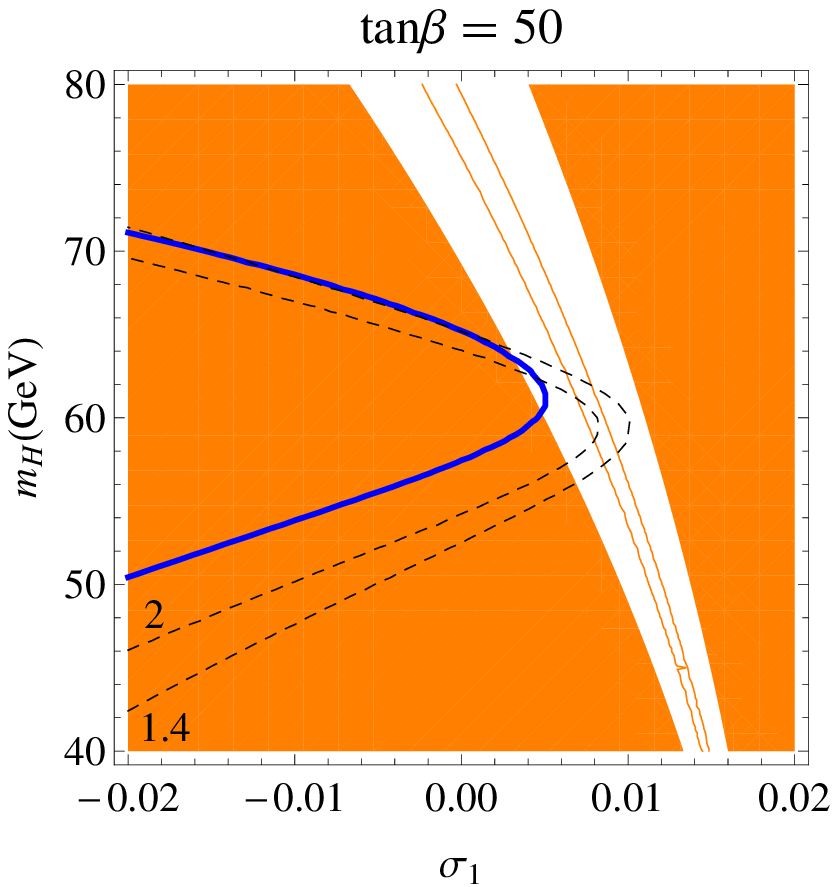}
  \end{center}
\end{minipage}
\caption{
 Same as Fig.~\ref{Fig:0240} but for $m_{\rm DM}=30$ GeV. 
}
\label{Fig:0230}
\end{figure}

A pair of scalar dark matter particles with mass $m_\phi=31-40$ GeV dominantly annihilates into $b \bar{b}$ 
   through the $s$-channel exchange of the Higgs bosons ($h$ and $H$). 
The cross section is enhanced by $H$-boson exchange when $m_H \sim 2 m_\phi$. 
We evaluate the cross section as a function of the coupling $\sigma_1$ and 
   the non-SM-like Higgs boson mass $m_H$ with fixed values for $\sigma_2$. 
Figures~\ref{Fig:0240} and \ref{Fig:0230} show the results for $m_{\phi} = 40$ and $30$ GeV
 for $\tan\beta=50, 40$, and $30$ with $\sigma_2=0.02$,
 which corresponds to a relatively large invisible decay rate of $h$, ${\rm BR(h \to \phi \phi})\sim 0.2$. 
Along the thick blue line, the observed DM relic density $\Omega h^2=0.1$ is reproduced, 
   while the two dashed lines correspond to the parameters that yield the present DM annihilation cross sections, 
   $(\sigma v)_0=1.4$ and 2.0, respectively, in units of $ 10^{-26}$ cm$^3$/s . 
The parameters in the overlapping region of the thick solid line and
 the region between the two dashed lines
 will both fit the gamma ray excess and reproduce the observed relic abundance.   
We also calculate the cross section of DM elastic scattering off nuclei, which is constrained 
   by the null results of the current direct DM detection experiments. 
The shaded regions are excluded by the LUX (2014) experiment~\cite{LUX:2013}, 
   and the expected sensitivity in future direct DM search experiments (
   such as the XENON1T experiment~\cite{XENON1T}) is depicted by two thin lines.
We find that there is no solution for $\tan\beta\lesssim 30$.

\begin{figure}[htbp]
 \begin{minipage}{0.32\hsize}
  \begin{center}
\includegraphics[width=50mm]{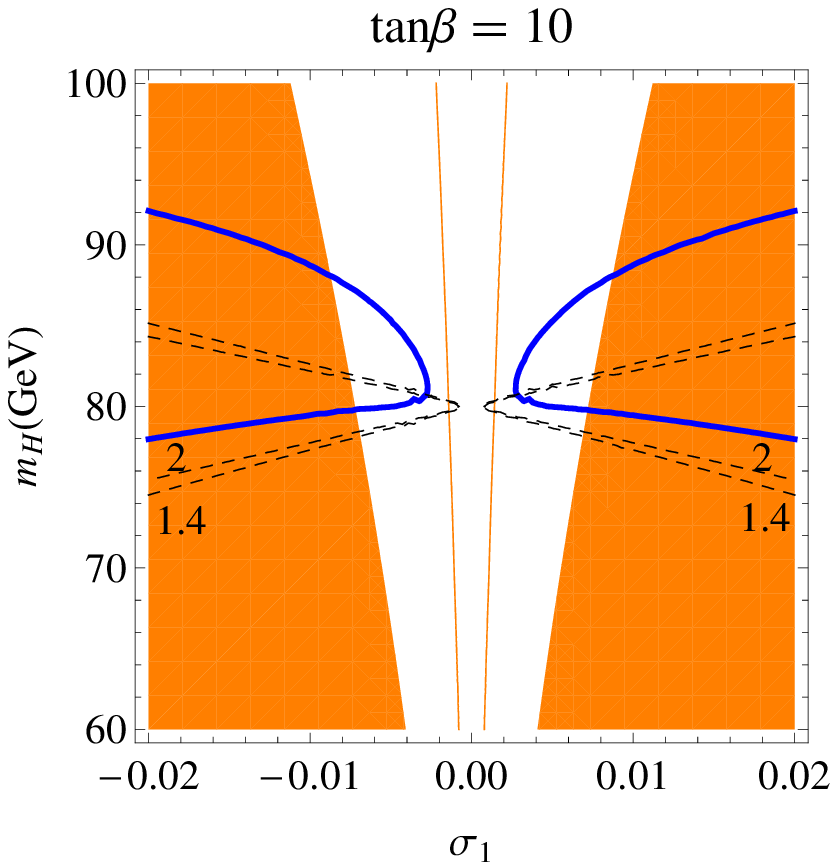}
  \end{center}
\end{minipage}
\hfill
 \begin{minipage}{0.32\hsize} 
  \begin{center}
\includegraphics[width=50mm]{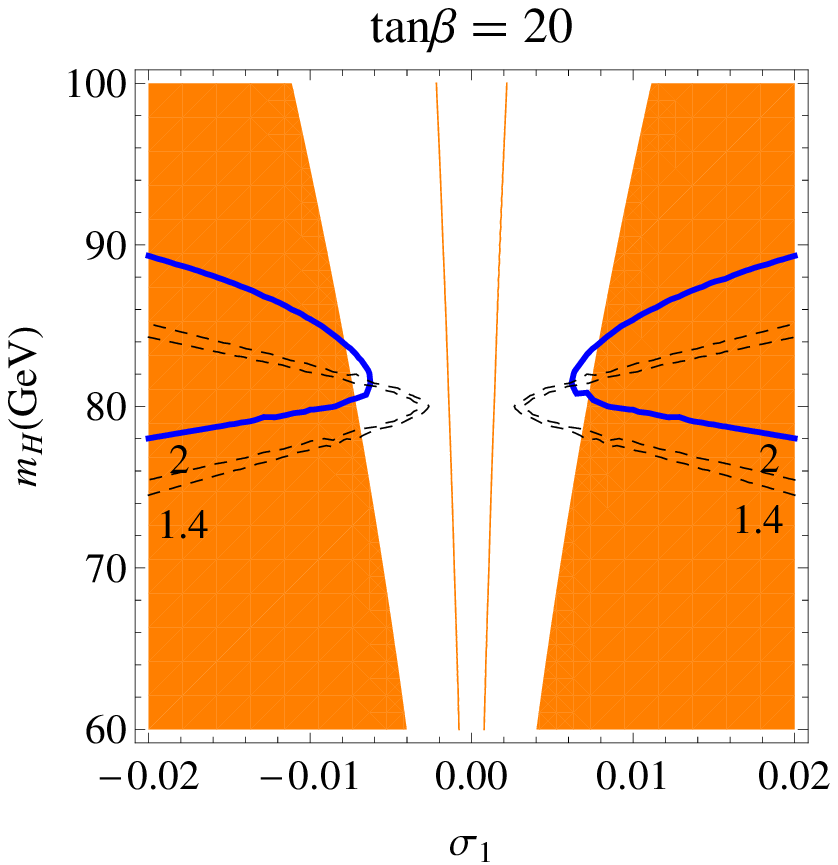}
  \end{center}
\end{minipage}
\hfill
 \begin{minipage}{0.32\hsize} 
  \begin{center}
\includegraphics[width=50mm]{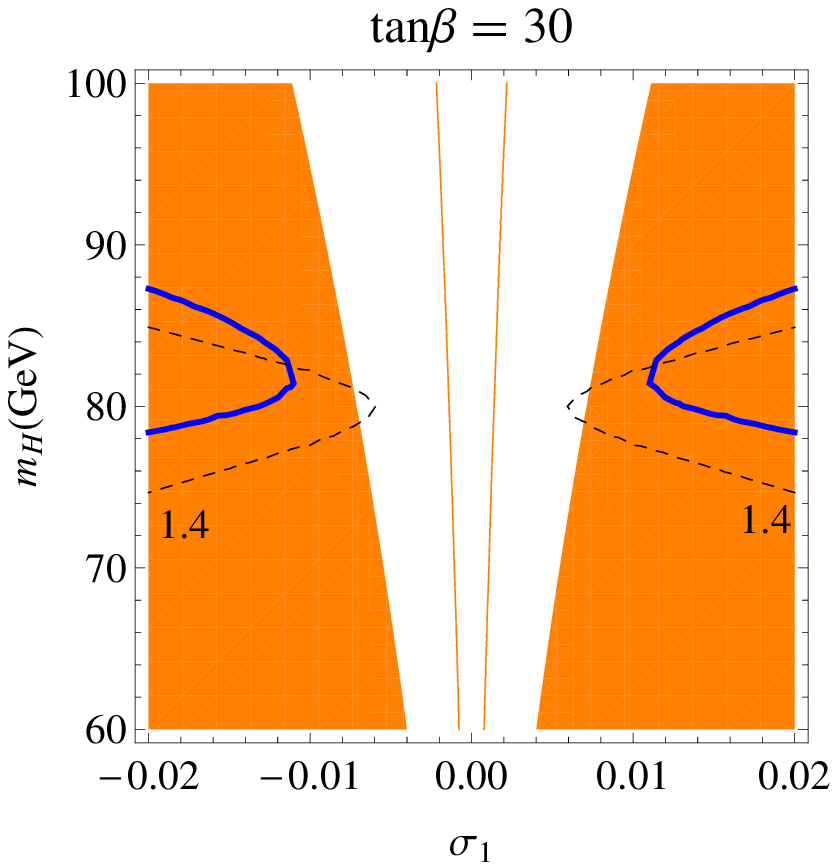}
  \end{center}
\end{minipage}
\caption{
 Same as Fig.~\ref{Fig:0240} but for $\sigma_2=0$.
$\tan\beta$ is taken to be $10, 20$, and $30$ from left to right.
}
\label{Fig:0040}
\end{figure}

Figure~\ref{Fig:0040} shows the results for $\sigma_2=0$, 
   which corresponds to a negligible invisible decay rate of $h$. 
We see that the results for a large $\tan\beta \gtrsim 30$ is already excluded 
   by the LUX experiment. 
The currently allowed parameter region will be covered by future direct DM detection experiments. 

\begin{figure}[h,t]
\begin{center}
\epsfig{file=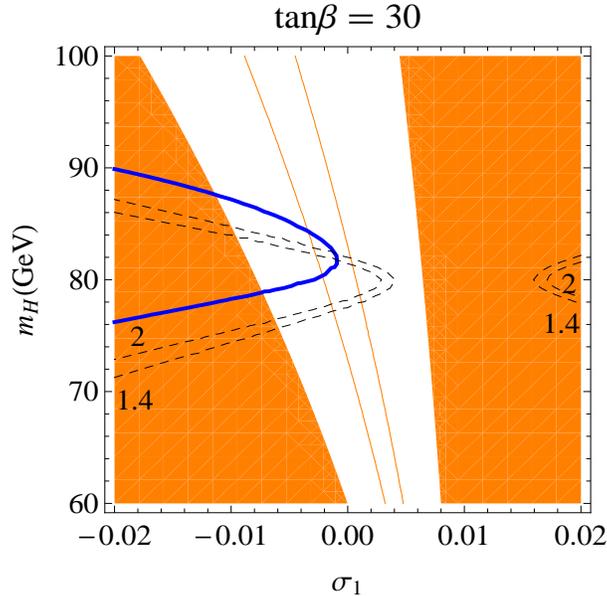, width=8cm,height=8cm,angle=0}
\end{center}
\caption{ Same as Fig.~\ref{Fig:0240} but for $\tan\beta=30$ and $\sigma_2=0.01$.
 }
\label{Fig:0140}
\end{figure}

From the above results, we find a correlation between $\sigma_2$ (in other words, the invisible decay
 rate of the SM-like Higgs boson $h$) and $\tan\beta$ in order to find viable parameter regions;
namely, a larger value of  $\sigma_2$ requires a larger $\tan\beta$ value. 
In fact, with $\sigma_2=0.02$, the large present DM annihilation cross section is obtained
 for $40 \lesssim \tan\beta \lesssim 50$, as seen in Figs.~\ref{Fig:0240} and \ref{Fig:0230}.
On the other hand, 
 for $\sigma_2=0.00$, a smaller $\tan\beta \lesssim 20$ is needed
 to avoid the direct DM search bound, as shown in Fig.~\ref{Fig:0040}.
We see in Fig.~\ref{Fig:0140} that the $\tan\beta=30$ case becomes available
 for a middle size of $\sigma_2$, say, $\sigma_2 \sim 0.01$.

\section{Summary}
Motivated by the gamma-ray excess from the Galactic Center and its interpretation 
  as annihilating dark matter particles,  we proposed a Higgs-portal DM scenario 
  in the context of the type-II two-Higgs-doublet model. 
This model can account for the gamma-ray excess through its main annihilation mode 
  into a pair of $b {\bar b}$ quarks via the $s$-channel exchange of the non-SM-like 
  Higgs boson with a type-II Yukawa coupling. 
We have identified the model parameter region that can explain the gamma-ray excess 
  and satisfies the phenomenological constraints on the relic dark matter abundance 
  and elastic scattering cross section of the DM particle off nuclei, as well as 
  the invisible decay rate of the SM-like Higgs boson into a pair of DM particles. 
Most of the identified parameter region can be tested by future direct dark matter detection experiments. 
In addition, the search for the invisible decay process of the SM-like Higgs boson and 
   the non-SM-like Higgs boson at future collider experiments is complementary to direct DM searches.


\section*{Acknowledgments}
This work is supported in part by 
 the DOE Grant No. DE-FG02-10ER41714 (N.O) 
and by the Grant-in-Aid for Scientific Research 
 on Innovative Areas No.~26105514 from
 the Ministry of Education, Culture, Sports, Science and Technology in Japan (O.S).
%


\appendix





\begin{thebibliography}{99}


\bibitem{Goodenough:2009gk} 
  L.~Goodenough and D.~Hooper,
  arXiv:0910.2998 [hep-ph].

\bibitem{Hooper:2010mq} 
  D.~Hooper and L.~Goodenough,
  Phys.\ Lett.\ B {\bf 697}, 412 (2011).

\bibitem{Hooper:2011ti} 
  D.~Hooper and T.~Linden,
  Phys.\ Rev.\ D {\bf 84}, 123005 (2011).

\bibitem{Abazajian:2012pn} 
  K.~N.~Abazajian and M.~Kaplinghat,
  Phys.\ Rev.\ D {\bf 86}, 083511 (2012).


\bibitem{Abazajian:2014fta} 
  K.~N.~Abazajian, N.~Canac, S.~Horiuchi and M.~Kaplinghat,
  Phys.\ Rev.\ D {\bf 90}, 023526 (2014).
  
\bibitem{Su:2010qj} 
  M.~Su, T.~R.~Slatyer and D.~P.~Finkbeiner,
  Astrophys.\ J.\  {\bf 724}, 1044 (2010).

\bibitem{Hooper:2013rwa} 
  D.~Hooper and T.~R.~Slatyer,
  Phys.\ Dark Univ.\  {\bf 2}, 118 (2013).

\bibitem{Huang:2013pda} 
  W.~-C.~Huang, A.~Urbano and W.~Xue,
  arXiv:1307.6862 [hep-ph].

\bibitem{Berlin:2013dva} 
  A.~Berlin and D.~Hooper,
  Phys.\ Rev.\ D {\bf 89}, 016014 (2014).


\bibitem{Hagiwara:2013qya} 
  K.~Hagiwara, S.~Mukhopadhyay and J.~Nakamura,
  Phys.\ Rev.\ D {\bf 89}, 015023 (2014).

\bibitem{Kyae:2013qna} 
  B.~Kyae and J.~-C.~Park,
  Phys.\ Lett.\ B {\bf 732}, 373 (2014).
  
\bibitem{Okada:2013bna} 
  N.~Okada and O.~Seto,
  Phys.\ Rev.\ D {\bf 89}, 043525 (2014).

\bibitem{Huang:2013apa} 
  W.~-C.~Huang, A.~Urbano and W.~Xue,
  JCAP {\bf 1404}, 020 (2014).
  
\bibitem{Modak:2013jya} 
  K.~P.~Modak, D.~Majumdar and S.~Rakshit,
  arXiv:1312.7488 [hep-ph].
  
\bibitem{Daylan:2014rsa} 
  T.~Daylan, D.~P.~Finkbeiner, D.~Hooper, T.~Linden, S.~K.~N.~Portillo, N.~L.~Rodd and T.~R.~Slatyer,
  arXiv:1402.6703 [astro-ph.HE].

\bibitem{Bringmann:2014lpa} 
  T.~Bringmann, M.~Vollmann and C.~Weniger,
  arXiv:1406.6027 [astro-ph.HE].

  
\bibitem{Lacroix:2014eea} 
  T.~Lacroix, C.~Boehm and J.~Silk,
  Phys.\ Rev.\ D {\bf 90}, 043508 (2014).
  
\bibitem{Yuan:2014rca} 
  Q.~Yuan and B.~Zhang,
  JHEAp {\bf 3-4}, 1 (2014).

\bibitem{Carlson:2014cwa} 
  E.~Carlson and S.~Profumo,
  Phys.\ Rev.\ D {\bf 90}, 023015 (2014).

\bibitem{Berlin:2014pya} 
  A.~Berlin, P.~Gratia, D.~Hooper and S.~D.~McDermott,
  Phys.\ Rev.\ D {\bf 90}, 015032 (2014).
\bibitem{Cheung:2014lqa} 
  C.~Cheung, M.~Papucci, D.~Sanford, N.~R.~Shah and K.~M.~Zurek,
  arXiv:1406.6372 [hep-ph].
\bibitem{Huang:2014cla} 
  J.~Huang, T.~Liu, L.~T.~Wang and F.~Yu,
  arXiv:1407.0038 [hep-ph].

\bibitem{Cerdeno:2014cda} 
  D.~G.~Cerdeno, M.~Peiro and S.~Robles,
  JCAP {\bf 1408}, 005 (2014).

\bibitem{Cerdeno:2008ep} 
  D.~G.~Cerdeno, C.~Munoz and O.~Seto,
  Phys.\ Rev.\ D {\bf 79}, 023510 (2009); \\  
  D.~G.~Cerdeno and O.~Seto,
  JCAP {\bf 0908}, 032 (2009).

\bibitem{Ghosh:2014pwa} 
  D.~K.~Ghosh, S.~Mondal and I.~Saha,
  arXiv:1405.0206 [hep-ph].


\bibitem{Boehm:2014hva} 
  C.~Boehm, M.~J.~Dolan, C.~McCabe, M.~Spannowsky and C.~J.~Wallace,
  JCAP {\bf 1405}, 009 (2014).
  
\bibitem{Agrawal:2014una} 
  P.~Agrawal, B.~Batell, D.~Hooper and T.~Lin,
  Phys.\ Rev.\ D {\bf 90}, 063512 (2014).

\bibitem{Izaguirre:2014vva} 
  E.~Izaguirre, G.~Krnjaic and B.~Shuve,
  Phys.\ Rev.\ D {\bf 90}, 055002 (2014).
  
\bibitem{Ipek:2014gua} 
  S.~Ipek, D.~McKeen and A.~E.~Nelson,
  Phys.\ Rev.\ D {\bf 90}, 055021 (2014).

\bibitem{Ko:2014gha} 
  P.~Ko, W.~-I.~Park and Y.~Tang,
  JCAP {\bf 1409}, 013 (2014).
 
\bibitem{Abdullah:2014lla} 
  M.~Abdullah, A.~DiFranzo, A.~Rajaraman, T.~M.~P.~Tait, P.~Tanedo and A.~M.~Wijangco,
  Phys.\ Rev.\ D {\bf 90}, 035004 (2014).
  
\bibitem{Martin:2014sxa} 
  A.~Martin, J.~Shelton and J.~Unwin,
  arXiv:1405.0272 [hep-ph].
  
\bibitem{Basak:2014sza}  
  T.~Basak and T.~Mondal,
  arXiv:1405.4877 [hep-ph].

\bibitem{Cline:2014dwa} 
  J.~M.~Cline, G.~Dupuis, Z.~Liu and W.~Xue,
  JHEP {\bf 1408}, 131 (2014).
  
\bibitem{Wang:2014elb}  
  L.~Wang,
  arXiv:1406.3598 [hep-ph].

\bibitem{Arina:2014yna} 
  C.~Arina, E.~Del Nobile and P.~Panci,
  arXiv:1406.5542 [hep-ph].

\bibitem{Ko:2014loa} 
  P.~Ko and Y.~Tang,
  arXiv:1407.5492 [hep-ph].

\bibitem{McDonald:1993ex}
  J.~McDonald,
  Phys.\ Rev.\  D {\bf 50}, 3637 (1994).
\bibitem{Burgess:2000yq}
  C.~P.~Burgess, M.~Pospelov and T.~ter Veldhuis,
  Nucl.\ Phys.\  B {\bf 619}, 709 (2001).
\bibitem{Davoudiasl:2004be}
  H.~Davoudiasl, R.~Kitano, T.~Li and H.~Murayama,
  Phys.\ Lett.\  B {\bf 609}, 117 (2005).
\bibitem{Kikuchi:2007az}
  T.~Kikuchi and N.~Okada,
  Phys.\ Lett.\  B {\bf 665}, 186 (2008). 

\bibitem{KMNO}
S.~Kanemura, S.~Matsumoto, T.~Nabeshima and N.~Okada,
  Phys.\ Rev.\ D {\bf 82}, 055026 (2010).

\bibitem{deSimone:2014pda} 
  A.~De Simone, G.~F.~Giudice and A.~Strumia,
  JHEP {\bf 1406}, 081 (2014).
  


\bibitem{Aoki:2009pf} 
  M.~Aoki, S.~Kanemura and O.~Seto,
  Phys.\ Lett.\ B {\bf 685}, 313 (2010).
\bibitem{Goh:2009wg} 
  H.~-S.~Goh, L.~J.~Hall and P.~Kumar,
  JHEP {\bf 0905}, 097 (2009).
\bibitem{Cai:2011kb} 
  Y.~Cai, X.~-G.~He and B.~Ren,
  Phys.\ Rev.\ D {\bf 83}, 083524 (2011).
\bibitem{Boucenna:2011hy} 
  M.~S.~Boucenna and S.~Profumo,
  Phys.\ Rev.\ D {\bf 84}, 055011 (2011).

\bibitem{Aoki:2008av} 
  M.~Aoki, S.~Kanemura and O.~Seto,
  Phys.\ Rev.\ Lett.\  {\bf 102}, 051805 (2009); 
  Phys.\ Rev.\ D {\bf 80}, 033007 (2009).
  

\bibitem{LEP2}
  R.~Barate {\it et al.}  [LEP Working Group for Higgs Boson Searches and ALEPH and DELPHI and L3 and OPAL Collaborations],
  Phys.\ Lett.\ B {\bf 565}, 61 (2003). 


\bibitem{Hinv}
  G.~Belanger, B.~Dumont, U.~Ellwanger, J.~F.~Gunion and S.~Kraml, 
  Phys.\ Lett.\ B {\bf 723}, 340 (2013).

\bibitem{KolbTurner}
  E.~W.~Kolb and M.~S.~Turner, {\it The Early Universe} (Addison-Wesley, Reading, MA, 1990).

\bibitem{WMAP} 
  G.~Hinshaw {\it et al.}  [WMAP Collaboration],
  Astrophys.\ J.\ Suppl.\ Ser.\  {\bf 208}, 19 (2013).

\bibitem{Planck} 
  P.~A.~R.~Ade {\it et al.}  [Planck Collaboration],
  arXiv:1303.5076 [astro-ph.CO].



\bibitem{LUX:2013} 
  D.~S.~Akerib {\it et al.}  [LUX Collaboration],
  Phys.\ Rev.\ Lett.\  {\bf 112}, 091303 (2014).

\bibitem{XENON1T}
 E.~Aprile [XENON1T Collaboration],
  Springer Proc.\ Phys.\ {\bf 148}, 93 (2013).



\end{thebibliography}
\end{document}